\documentstyle[twoside,fleqn,espcrc2,epsfig]{article}
\newcommand{\csw}{c_{\mbox{sw}}}
\newcommand{\kcrit}{\kappa_{\rm crit}}
\newcommand{\plus}{\makebox[30pt][c]{$+$}}
\newcommand{\minus}{\makebox[30pt][c]{$-$}}

\newcommand{\er}[2]{
  \raisebox{0.08em}{\scriptsize {$\hspace{-0.3em}\begin{array}{@{}l@{}}
        \plus\makebox[0.15em][r]{#1} \\[-0.12em] 
        \minus\makebox[0.15em][r]{#2} 
      \end{array}$}}}
\title{Calculation of $f_B$ and the ``Isgur-Wise Function'' using a
  non-perturbatively improved fermion action}

\author{UKQCD Collaboration, presented by D.G.~Richards\address{Dept.\ 
    of Physics and Astronomy, University of Edinburgh, Mayfield Road,
    Edinburgh EH9 3JZ, UK}}

\begin{document}
\begin{abstract}
We present a calculation of $f_B$ and of the form factors for
the semi-leptonic decay $\overline{B} \rightarrow D l
\bar{\nu}$ in the quenched approximation to QCD.  Results are
generated on lattices at $\beta = 6.2$, using an $O(a)$-improved
fermion action, with the clover coefficient determined
non-perturbatively.
\end{abstract}
\maketitle
\section{Introduction}
A knowledge of the hadronic matrix elements for the decays of mesons
containing heavy quarks is essential to the
experimental determination of elements of the CKM matrix involving
the heavy quarks.  In this contribution, we concentrate on the hadronic
matrix elements for the semi-leptonic decay $\bar{B} \rightarrow D l
\bar{\nu}$, and also discuss $f_B$.  The calculations of
the $D \rightarrow K$ and $D \rightarrow \pi$ matrix elements are
contained in the talk of Chris Maynard\cite{chrism}.  An important
element of this programme is the use of a ``clover'' fermion action
with the clover coefficient, $\csw$, determined non-perturbatively.
Thus we expect a substantial reduction in discretisation errors
compared to earlier calculations using the tree-level value of
$\csw$.

\section{Simulation details}
The calculation is performed in the quenched approximation on an
ensemble of 216 $24^3 \times 48$ lattices at $\beta = 6.2$, generated
using the Wilson gauge action.  In the calculation of the the
semi-leptonic decay matrix element, we used four values of the final
heavy-quark mass, corresponding to $\kappa_h = 0.1200, 0.1233, 0.1266$
and $0.1299$, and two values of the initial heavy-quark mass,
corresponding to $\kappa_h = 0.1200$ and $0.1266$; the charm-quark mass
corresponds roughly to $\kappa = 0.1233$.  Further details of the
calculation are contained in ref.~\cite{chrism}.

In the non-perturbatively improved scheme, the axial vector and
pseudoscalar currents mix:
\begin{eqnarray}
A_\mu^{\mbox{\scriptsize R}}
&=& Z_A\bigl(1 + b_A am_q\bigr)\nonumber\\
& & \times
    \Bigl[A_\mu^{\mbox{\scriptsize latt}}
    + c_A \frac{a}{2}(\partial_\mu^\ast + \partial_\mu)
    P^{\mbox{\scriptsize latt}}\Bigr]
\end{eqnarray}
where
\begin{equation}
m_q = \frac{1}{2}(\frac{1}{\kappa} - \frac{1}{\kcrit}).
\end{equation}
We take the ALPHA Collaboration's numerical values\cite{alpha_np} for
the overall renormalisation factor, $Z_A$ and for $c_A$, and the
one-loop calculation of $b_A$.  Note that this form is modified for
the case of a current constructed of non-degenerate quarks through the
addition of a term linear in the difference of the quark masses, with
coefficient $b_A'$; this coefficient not been computed.

The renormalisation of the vector current proceeds likewise:
\begin{eqnarray}
V_{\mu}^{\rm R} & = & Z_V (1 + b_V a m_q) \times \nonumber\\
& & \left[ V_{\mu}^{\rm latt} +
  c_V \frac{a}{2} (\partial_{\nu}^{*} + \partial_{\nu})
  \Sigma_{\mu\nu}^{\rm latt}
  \right],\label{eq:alpha_zv}
\end{eqnarray}
where $Z_V$, $b_V$ and $c_V$ are all known non-perturbatively.

\section{Pseudoscalar decay constant}
We obtain $f_P/f_{\pi}$ at each value of $\kappa_l$ and $\kappa_h$,
where $P$ is the heavy-light pseudoscalar, and $f_{\pi}$ is the decay
constant for a light meson constructed from degenerate quarks of mass
corresponding to $\kappa_l$.  We then extrapolate
$\kappa_l$ to $\kcrit$ at fixed $\kappa_h$.  Note that, whilst $Z_A$
cancels between the denominator and numerator, an effective relative
renormalisation enters through $b_A$.

\begin{figure}
\epsfxsize=200pt \epsfbox{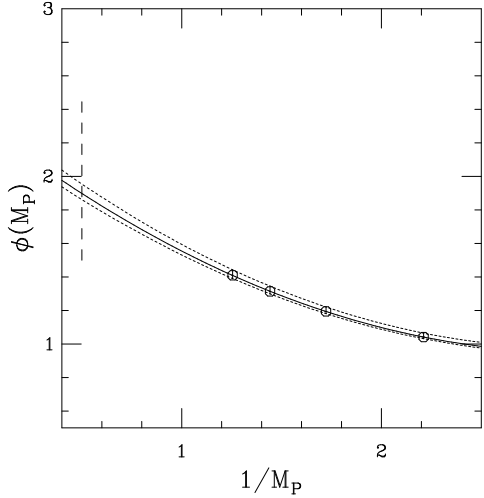}
\caption{The ratio $\phi(M_p)$ is shown at $\kappa_l = \kappa_{\rm
    crit}$ for each value of $\kappa_h$.  The solid line is a quadratic fit
  to the data, and the dashed line corresponds to the extrapolation to
  $M_B$.}\label{fig:fb}
\end{figure}

To investigate the heavy-mass dependence, we construct the scaling
  quantity
\begin{equation}
\Phi \equiv (\alpha_s(M_P)/\alpha_s(M_B))^{2/\beta_0}
(f_P/f_{\pi})\sqrt{M_P},
\end{equation}
where we take $N_f = 0$, and perform a quadratic fit in $1/M_P$, as
shown in Figure~\ref{fig:fb}.
Setting $a^{-1} = 2.64\, \mbox{GeV}$, the value obtained from
$m_{\rho}$, we find
\begin{eqnarray}
f_D/f_{\pi} & = & 1.19\er{3}{1}\nonumber \\
f_B/f_{\pi} & = & 1.34\er{4}{3},
\end{eqnarray}
where the quoted errors are purely statistical, and hence
\begin{eqnarray}
f_D & = & 190 \er{5}{2}\quad \mbox{MeV}\nonumber\\
f_B & = & 176 \er{5}{4}\quad \mbox{MeV}.
\end{eqnarray}

\section{Semi-leptonic decays and the
  Isgur-Wise function}
The form factors for the semi-leptonic decay $\overline{P} \rightarrow
P' l \bar{\nu}$, where $P^{(')}$ is a heavy-light pseudoscalar meson,
may be parametrised as
\begin{eqnarray}
\lefteqn{\frac{\langle P'({\bf p}') | \bar{Q}' \gamma_{\mu} Q | P({\bf
    p})\rangle}{\sqrt{M_P M_{P'}}} =
 (v + v')_{\mu} h^{+}(\omega; m_Q,
    m_{Q'})}\nonumber \\
& &  + (v - v')_{\mu} h^{-}(\omega; m_Q, m_{Q'}).
\end{eqnarray}
Here $v^{(')}$ is the four velocity of the initial (final) meson, and
$\omega = v \cdot v'$.  In the Heavy Quark Effective Theory (HQET),
the form factors display an additional spin-flavour symmetry, and are
related to a universal ``Isgur-Wise'' function $\xi(\omega)$:
\begin{eqnarray}
\lefteqn{h^i(\omega; m_Q, m_{Q'}) = \xi(\omega) \times}\nonumber\\
& &(\alpha^i + \beta^{i}(\omega; m_Q, m_{Q'})
+ \gamma^i (\omega; m_Q, m_{Q'}))
\end{eqnarray}
where $\alpha^{+} = 1$, $\alpha^{-} = 0$, and $\beta^i$ and $\gamma^i$
represent the radiative and power corrections respectively.  Note that
$\xi$ is normalised: $\xi(\omega = 1) = 1$.

\begin{figure}
\epsfxsize=200pt \epsfbox{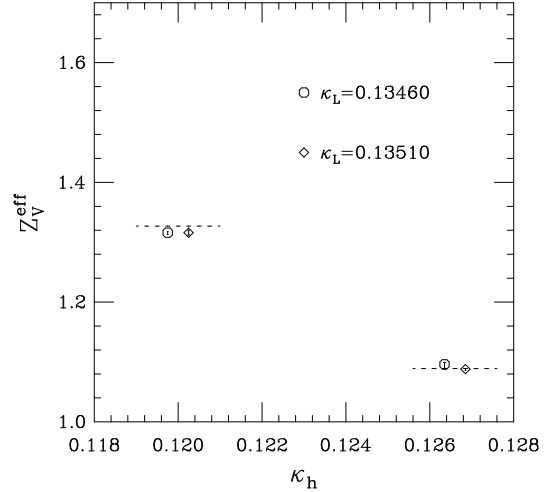}
\caption{$Z_V^{\rm eff}$ is shown for two values of the light-quark
  mass and for two values of the heavy-quark mass.  The dashed lines
  represent the predictions using the non-perturbative values of
  eqn.~\ref{eq:alpha_zv}}\label{fig:zv}
\end{figure}
For the case of degenerate transitions at zero momentum transfer,
we can obtain a direct measurement of the
effective renormalisation constant
\begin{equation}
Z_V^{\rm eff} = Z_V ( 1 + b_V a m_Q),
\end{equation}
and compare with the value using the non-perturbative parameters of
eqn.~\ref{eq:alpha_zv}, as shown in Figure~\ref{fig:zv}.  The
agreement is striking, and perhaps surprising since the NP
prescription only removes ${\cal O}(am_Q)$ errors.

\begin{figure}
\epsfxsize=200pt \epsfbox{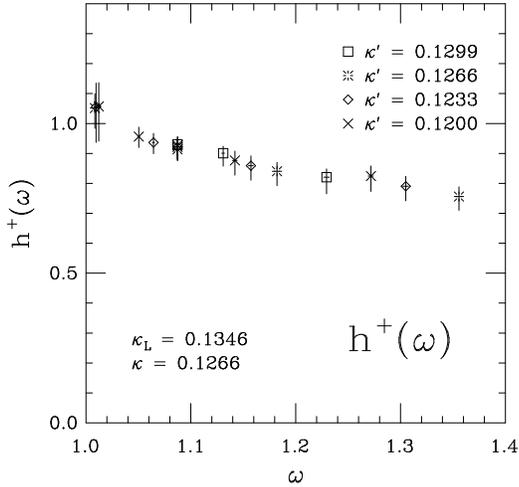}
\caption{$h^{+}(\omega)$ is shown for fixed light-quark and initial
  heavy-quark masses.}\label{fig:hplus}
\end{figure}

\begin{figure}
\epsfxsize=200pt\epsfbox{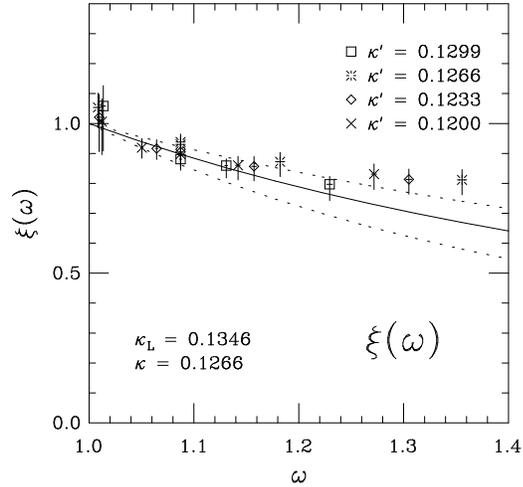}
\caption{$\xi(\omega) \equiv h^{+}(\omega)/(1 + \beta^{+}(\omega))$ is
  shown for the same parameter values as in Figure~\ref{fig:hplus}.
  The solid line is a fit to the data for $\kappa' =
  0.1233$, as described in the text.}\label{fig:wisgur}
\end{figure}

In Figure~\ref{fig:hplus}, we show the form factor $h^{+}(\omega)$ for
fixed value of the light-quark mass, close to the strange, for fixed
initial heavy-quark mass, and for four values of the final heavy-quark
mass; as expected, $h^{-}$ is very small.
We compute $\beta^{+}$using Neubert's short-distance expansion of the
currents\cite{neubert92}, and ``define'' the Isgur-Wise function
through
\begin{equation}
\xi(\omega) = \frac{h^{+}(\omega)}{1 + \beta^{+}(\omega)}.
\end{equation}
The corrected form factor is shown in Figure~\ref{fig:wisgur},
together with
a one-parameter fit for $\omega < 1.2$ to
\[
\xi(\omega) = \frac{2}{1 + \omega} \exp (-(2 \rho^2 - 1)(\omega -
1)/(\omega + 1)).
\]

The use of the NP prescription
has enabled a far more satisfactory treatment of discretisation errors
than the calculation using the tree-level-improved SW action\cite{ukqcd}.
It remains to obtain a quantitative estimate of the power corrections,
and a determination of the remaining form factors.

\section{Acknowledgements}
I acknowledge PPARC through the support of an Advanced Fellowship.
T3D time was funded through EPSRC grant GR/K41663.


\begin{thebibliography}{9}
\bibitem{chrism} UKQCD Collaboration, presented by Chris Maynard,
  these proceedings.
\bibitem{alpha_np} M.~L\"{u}scher \textit{et al.}, Nucl.\ Phys.\
  {\bfseries B478} (1996) 365; Nucl.\ Phys.\ {\bfseries B491} (1997) 344.
\bibitem{neubert92} M.~Neubert, Phys.~Rev.~{\bf D46} (1992) 2212.
\bibitem{ukqcd} UKQCD, S.P.~Booth \textit{et al.}, Phys.\ Rev.\ Lett.\
  {\bfseries 72} (1994) 462; Phys.\ Rev.\ {\bf D51} (1995) 4905.
\end{thebibliography}
\end{document}